\documentclass[aps,twocolumn,prd,showpacs]{revtex4}
\usepackage{amsmath}
\usepackage{graphicx}
\usepackage{amsmath}
\usepackage{amssymb}
\usepackage{latexsym}
\usepackage{color}
\newcommand{\de}{\mathrm{d}}

\newcommand{\p}{\partial}

\begin{document}

\title{Cosmological perturbations of a perfect fluid and non-commuting variables}

\author{Antonio De Felice}
\affiliation{Centre for Particle Physics and Phenomenology (CP3),
  Universit\'e catholique de Louvain, Chemin du Cyclotron 2, B-1348
  Louvain-la-Neuve, Belgium}

\author{Jean-Marc G\'erard}
\affiliation{Centre for Particle Physics and Phenomenology (CP3),
  Universit\'e catholique de Louvain, Chemin du Cyclotron 2, B-1348
  Louvain-la-Neuve, Belgium}

\author{Teruaki Suyama}
\affiliation{Centre for Particle Physics and Phenomenology (CP3),
  Universit\'e catholique de Louvain, Chemin du Cyclotron 2, B-1348
  Louvain-la-Neuve, Belgium}

\date{\today}

\begin{abstract}
  We describe the linear cosmological perturbations of a perfect fluid
  at the level of an action, providing thus an alternative to the
  standard approach based only on the equations of motion. This action is
  suited not only to perfect fluids with a barotropic equation of
  state, but also to those for which the pressure depends on two
  thermodynamical variables. By quantizing the system we find that
  1)~some perturbation fields exhibit a non-commutativity quite
  analogous to the one observed for a charged particle moving in a
  strong magnetic field, 2)~local curvature and pressure perturbations
  cannot be measured simultaneously, 3)~ghosts
  appear if the null energy condition is violated.
\end{abstract}
\maketitle

\section{Introduction}

The theory of cosmological perturbations (TCP) for a perfect fluid has
always been an important issue in cosmology.  It enables us to
understand how small fluctuations seeded in the early universe eventually evolved
into the present large scale structure.  Also, TCP has been extremely
useful to put constraints on various cosmological models.

TCP for a perfect fluid has been developed and studied at the level of
the basic equations of motion, i.e., the Einstein equations of general
relativity (GR) and the energy-momentum conservation law \cite{Kodama:1985bj,Mukhanov:1990me}. 
Yet, TCP for a perfect fluid can be also studied at the level of the action.
Although these two approaches are classically equivalent,
the latter gives the following advantage.
In TCP, one first has to perturb all the fields
appearing in the equations of motion or in the action, such as the metric 
components and the energy density.
However, as is well known, not all the perturbation fields are dynamical.
Actually, GR with a perfect fluid has only one dynamical field for 
the scalar-type perturbation.
But an identification of this field as well as a derivation of its closed evolution equation by means
of the equations of motion alone are not straightforward.

The situation becomes worse when going to extended gravity models.
For illustration, $f(R,G)$ theories ($R$ being the Ricci scalar, and $G$ the Gauss-Bonnet term) 
with a perfect fluid involve two dynamical fields for the scalar-tensor perturbation.
In such theories, the usual approach based on the equations of motion requires a rather strong intuition 
because the closed evolution equations for those dynamical fields have to be extracted from rather 
complicated coupled differential equations.
On the other hand, the action approach advocated in this paper allows a straightforward identification of the
auxiliary fields just by checking the absence of any kinetic terms in the second order action.
Once the auxiliary fields are found, they can easily be eliminated through their trivial equations of
motion. What is then left is an action containing only the dynamical fields, from which we can derive the closed evolution equations.
In \cite{DeFelice:2009ak,DeFelice:2009wp},
we have explicitly checked that the action approach indeed works for $f(R,G)$
gravity models with no matter and with a scalar field, respectively.

%{\bf Furthermore, the background of
%perfect lfuids on a cosmological background have been quantized in the
%context of quantum cosmology, see e.g.\
%\cite{Pedram:2007ud,Pedram:2007er,Alvarenga:2001nm,Monerat:2005mx,Lemos:1995qu,Ivashchuk:1995uy,Peter:2006id,Brown:1989vb}.} 
In this paper, we want to describe first-order TCP for GR with a perfect fluid 
at the level of the action, 
in a way consistent with the principles of thermodynamics.
To this end, we use the action for a perfect fluid proposed by Schutz \cite{Schutz}
and do the quantization of the perturbations,
which might also be of some interest beyond a pure academic point of view.
Indeed, the quantization of the background universe with a perfect fluid has been 
discussed by many authors \cite{Pedram:2007ud,Pedram:2007er,Alvarenga:2001nm,Monerat:2005mx,Lemos:1995qu,Ivashchuk:1995uy,Peter:2006id,Brown:1989vb}.
%Such a method is quite useful when quantizing the fluid perturbation
%fields, which is the other main issue in this paper. 
But here, we prove that quantizing the perturbation fields leads to non-standard
commutation relations and, consequently, to unexpected effects upon the physical
properties of any perfect fluid in quantum cosmology. 
%{\bf We expect that such a quantization might lead to interesting cosmological signatures every time the quantization might be of importance. For example during reheating or just after it, for the study of quantum fluctuations also in the light of non-gaussianities, the TCP-action approach introduced here should be employed.}

A TCP action approach for fluids was first introduced in \cite{Garriga}
in the context of k-inflation. This approach was taken also in \cite{Boubekeur:2008kn}
to study non-linear cosmological perturbations in the matter dominated universe.
However, the fluid discussed there is the so-called scalar fluid whose energy-momentum
tensor is completely written in terms of a scalar field and its derivative.
By construction, the scalar fluid cannot have vector-type perturbations.
Although there is an exact correspondence between a perfect fluid and the scalar fluid
for the scalar-type perturbation at the linear order, 
it is no longer true for higher order perturbations because of the mixture of 
scalar and vector-type perturbations.
On the other hand, the Schutz's action we will use here is for a
perfect fluid. Therefore the action exactly describes the dynamics of a perfect fluid 
at any order.
As far as we know, this is the first time TCP is fully developed within the Schutz's action.
We believe our approach is suited for studying cosmology in extended gravity models.

%and hence could not generate vector perturbations, 
%so that it is physically different from a perfect fluid.

Before introducing the action for a perfect fluid, let us briefly
review the thermodynamics needed to describe it.  In this paper, we
consider a ``single'' fluid, that is, a fluid whose thermodynamical
quantities are completely determined by only two variables, e.g.\ the
chemical potential $\mu$ and the entropy per particle $s$
\cite{Misner}.  In this sense one first needs to give two equations of
state, $n=n(\mu,s)$ and $T=T(\mu,s)$, where $n$ is the number density
and $T$ is the temperature of the fluid.  Using then the first law of
thermodynamics, $\de p=n\de\mu-nT\de s$, one obtains the pressure as
$p=p(\mu,s)$.  Finally, the energy density is given by $\rho\equiv\mu
n-p$.  This is enough to describe the system thermodynamically.
Single fluids also satisfy particle number conservation, namely $N=n
V$ is a constant. The second law of thermodynamics imposes $\de(N
s)=N\de s\geq0$ such that $\de s=0$ at equilibrium.

A single perfect fluid is also defined through its stress-energy
tensor $T_{\mu\nu}=(\rho+p)u_\mu u_\nu+pg_{\mu\nu}$.  In a
Friedmann-Lema\^\i tre-Robertson-Walker (FLRW) background the
conservation of energy-momentum, $T^{\mu\nu}{}_{;\nu}=0$, implies that
$\dot\rho+3H(\rho+p)=0$ or, equivalently, $\de (\rho V)+p\,\de V=0$
since $V\propto a^3$ with $a$, the cosmological scale factor, and
$H\equiv\dot a/a$, the Hubble parameter. This, in turn, implies that
$\de N=0$ and $\de s=0$. In any FLRW universe we thus have
$na^3=N=\mathrm{constant}$ and $\dot s=0$.

\section{Action}

The action considered here has been introduced by Schutz \cite{Schutz} and is
defined as follows
\begin{equation}
  \label{eq:act1}
  S=\int d^4x\sqrt{-g}\left[\frac R{16\pi G}+p(\mu,s)\right] .
\end{equation}
Alternative functionals have been proposed, all being physically
equivalent as shown in \cite{sorkino}. We chose the version
(\ref{eq:act1}) as it was the most convenient for our purpose. The
four-velocity of the perfect fluid is defined via potentials \cite{Schutz}:
\begin{equation}
  \label{eq:vel1}
  u_{\nu}=\frac1\mu\, ( \p_{\nu}\ell+\theta\p_{\nu} s+A\p_{\nu} B)\, ,
\end{equation}
where $\ell$, $\theta$, $A$ and $B$ are all scalar fields. The
normalization for the four-velocity, $u^\nu u_\nu=-1$, gives $\mu$ in
terms of the other fields. The fundamental fields over which the
action (\ref{eq:act1}) will be varied are $g_{\mu\nu}$, $\ell$,
$\theta$, $s$, $A$, and $B$.
%{\bf It can be shown that equations obtained by the variations are Einstein equations 
%and the conservation law of the energy-momentum tensor \cite{Schutz}.}

Having chosen the Lagrangian for gravity to be the one of GR, we
recover $G_{\mu\nu}=8\pi G\, T_{\mu\nu}$ by varying with respect to
the metric field. Besides the conservation of particle number and
entropy already discussed, the other equations of motion derived from
Eq.\ (\ref{eq:act1}) are \cite{Schutz}:
\begin{equation}
\label{eq:therm}
u^\alpha\p_\alpha \theta=T,\quad u^\alpha \p_\alpha A=0,\quad u^\alpha\p_\alpha B=0.
\end{equation}
In a FLRW universe, $u_i=0$ and $u_0=-1$ such that the solutions to
Eq.~(\ref{eq:therm}) are simply
\begin{equation}
\label{eq:backA}
A=A(\vec x)\, ,\quad B=B(\vec x)\, ,\quad\theta=\int^t T(t')\de t'+\tilde\theta(\vec x)\, .
\end{equation}
There is a complete freedom for the functions $A$, $B$, and
$\tilde\theta$\ \footnote{ Since $u_\nu=(-1,\vec 0)$, we also have
  that $\ell=-\int^t \mu(t')\de t'+\tilde\ell$, and
  $\vec\nabla\tilde\ell=-A\vec\nabla B$, which implies that
  $\vec\nabla A\times\vec\nabla B=0$.}, any choice leading to the same
physical background. We will take advantage of this freedom to
simplify our study of the scalar and vector perturbations.
% Summarizing,
%\begin{itemize}
%\item perfect fluids are characterized by their normalized
%  4-velocity $u^\alpha$, their energy-momentum tensor,
%  $T^\alpha{}_\beta$, and by the two EOS needed to define them.
%\item Different time-independent backgrounds quantities $A$, $B$, and
%  $\tilde\theta$ lead to the same physical content, i.e.\ the same
%  $T^\alpha{}_\beta$, $u^\alpha$, $\mu(t)$ and $s(t)$.
%\end{itemize}

\section{Perturbations}

Once and for all, we work within a spatially flat FLRW universe. At
first order in perturbation theory we have $\delta u_0=\frac12\delta
g_{00}$ and%
\begin{equation}
\label{eq:Perto}
\delta u_i = \partial_i \left( \frac{\delta \ell + \theta \delta s + A \delta B}\mu \right)+
  \frac{W_i}\mu\, ,
\end{equation}
with
\begin{equation}
W_i\equiv B_{, i} \delta A -A_{,i} \delta B - \tilde{\theta}_{, i}\delta s\equiv
\partial_i w_s+\bar u_i\, .
  \label{eq:wi}
\end{equation}
Note that $W_i$ is gauge invariant since, following ref.\ \cite{Weinberg}, 
the perturbation fields transform respectively as  
\begin{alignat}{6}
  \delta\ell&\to\delta\ell
  +\mu\xi^0+A\,\partial_i B\,&&\xi^i,&&\\
  \delta s&\to\delta s,&&\delta \theta&&\to\delta \theta-T\,\xi^0-\partial_i\tilde\theta\,\xi^i,\\
  \delta A&\to\delta A-\partial_iA\,\xi^i,&&\delta B&&\to\delta B-\partial_iB\,\xi^i,
\label{eq:deltB}
\end{alignat}
under the gauge transformation $x^\alpha\to x^\alpha+\xi^\alpha$. In Eq.\ (\ref{eq:wi}) we have
decomposed $W_i$ into scalar ($w_s$) and divergence-less vector modes
($\bar u_i$). So, in general $W_i$ will generate both scalar and
vector perturbations. However, we can efficiently use the freedom of
choosing the time-independent background quantities $A$, $B$ and
$\tilde\theta$ given in Eq.\ (\ref{eq:backA}) to disentangle them. Any
such choice does not fix a gauge as no conditions are imposed on the
perturbation fields themselves.

\subsection{Scalar type perturbations}

Let us simply consider the choice
\begin{equation}
  \label{eq:choi}
  A=B=\tilde\theta=0\, ,
\end{equation}
to remove the vector perturbations arising from $W_i$.  Regarding the
metric, $\delta g_{00}$ and $\delta g_{0i}$ are auxiliary fields such
that the only scalar component which will be dynamical is the
curvature perturbation $\phi$ defined by $\delta g_{ij}=2a^2\phi\,\delta_{ij}$,
with $\phi \to \phi-H \xi^0$ under a gauge transformation.

We introduce the new quantity $v=\delta\ell+\theta(t)\delta s$ such that $\delta u_i=\partial_i \left( v/\mu \right)$.
Therefore, $v$ represents the velocity perturbation of a perfect fluid.
We then define two gauge invariant fields, $\Phi=\phi+Hv/\mu$ and
$\delta\bar\theta=\delta\theta+Tv/\mu$, to expand the action
(\ref{eq:act1}) at second order, in a gauge-independent way:
% {\bf Since we use gauge-invariant fields, the results will be valid and the same in any guage. For example $\Phi=\phi$ in the $v=0$ gauge, and so on. We find the action reduces to}
\begin{align}
  S_S&=\int \de t\de^3\vec x\left\{ \frac{a^3Q_S}2\left[\dot\Phi^2
-\frac {c_s^2}{a^2}(\vec\nabla\Phi)^2\right]
+C\delta s\dot\Phi-\frac D2 \delta s^2\right.\notag\\
&\qquad\left.{}
-E(\delta\bar\theta\dot{\delta s}
-\delta s\dot{\delta\bar\theta}
+\delta A\dot{\delta B}
-\delta B\dot{\delta A})\right\}.
\label{eq:act2}
\end{align}
The perturbation fields $\Phi$ and $\delta\bar\theta$ are related to the curvature
and temperature, respectively.
In the comoving gauge $v=0$ where a perfect fluid remains static, $\Phi=\phi$ and ${\delta\bar\theta}=\delta \theta$.
The coefficients for the kinetic terms are given by\
\footnote{For $c_s^2$ we used the fact that $\dot p= \left(\p
      p/\p\rho\right)_s\dot\rho+\left(\p p/\p
      s\right)_\rho\dot s$.}
\begin{alignat}{10}
Q_S&=\frac{\rho+p}{c_s^2 H^2}\, ,&\quad&&
c_s^2&\equiv\frac{\dot p}{\dot \rho}=
\left(\frac{\p p}{\p\rho}\right)_s ,&\quad&& \label{eq:Qs}
\end{alignat}
whereas the remaining coefficients are
\begin{alignat}{4}
C&=\frac{na^3}{H}\left[\mu\left(\frac{\p T}{\p\mu}\right)_{\!s}-T\right],
\quad&
E&=\frac{na^3}{2}\, ,
\end{alignat}
and
\begin{align}
D&=na^3\left[T\left(\frac{\p T}{\p\mu}\right)_{\!s}+\left(\frac{\p T}{\p s}\right)_{\!\mu}\right] .
\end{align}
The general solution for $\delta s$, $\delta A$, and $\delta B$ is
their initial values since Eq.\ (\ref{eq:act2}) forces them to be
time-independent. As a consequence, the non trivial equations of
motion are
\begin{align}
\label{eq:gPhi}
&  \frac{1}{a^3Q_S}\frac{\de}{\de t}(a^3Q_S\dot\Phi)-\frac{c_s^2}{a^2}\nabla^2\Phi=-\frac{\dot C}{a^3Q_S}\,\delta s\,,\\
& n a^3\dot{\delta\bar\theta}-D\delta s+C\dot\Phi=0\, .
\end{align}
These equations exactly coincide with those derived by perturbing the
Einstein equations and the conservation law for the entropy, as it should be.  In
general, $\Phi$ is sourced by $\delta s$.  For example, if the perfect
fluid is an ideal non-relativistic gas characterized by
$T=\frac25(\mu-m_0)$, $m_0$ being the mass of the particles, then
${\dot C} \neq 0$ and we have to solve two coupled equations to
know the time evolution of $\Phi$ and $\delta{\bar\theta}$.

However, if $T=f(s)\,\mu$, which is equivalent to having a barotropic
equation of state $p=p(\rho)$ \footnote{In this case, we obtain $\left(\p \mu/\p
    s\right)_\rho= T$ such that $(\p p/\p s)_\rho=n[\left(\p \mu/\p s\right)_\rho-T]=0$.}, then $C=0$.
(Note that both radiation and dust fulfill this condition, while a cosmological constant has vanishing
$Q_S$ so that no contribution for perturbations arises, as is well known).
In this case, the sign of $Q_S$ cannot be known from the usual approach based on the equations of motion alone. 
On the other hand, the action approach advertised here leads to an exact expression for $Q_S$, 
which will be used to avoid ghost degrees of freedom when quantizing the perturbations. 

We also conclude that the fields $\Phi$ completely decouples from $\delta s$ and propagates with a
sound speed $c_s$ if $C=0$ and $c_s^2 >0$.

\subsection{Vector type perturbations}

To arrive at the desired action via the shortest path, let us first
assume that all the perturbation variables propagate only in one
direction, say the $z$-direction.  This should be allowed, as we know
that perturbations with different wavenumber vectors do not mix in a
FLRW universe.  Once we obtain the action for this particular mode, we
can then easily infer the general action.

The vector contributions come only from the component $\bar u_i$ of
$W_i$ defined in Eq.\ (\ref{eq:wi}). It is not easy to extract $\bar
u_i$ from this equation since the functions $A,B$ and $\tilde{\theta}$
depend in general on the spatial coordinates. Yet, taking again advantage of
the freedom to select these background functions, we can make the
simplest choice that contains all the information needed for the
vector modes, namely
\begin{equation}
  A=\tilde\theta=0\,,\quad
B_{,i}=b_i\, ,
\end{equation}
where ${\vec b}=(b,0,0)$ is a constant vector orthogonal to the
$z$-direction. With this assumption, we have $w_s=0$ and $\bar u_i=b_i
\,\delta A(t,z)$ for $W_i$.

Regarding the vector perturbation of the metric, we follow again ref.\ \cite{Weinberg} 
and denote $\delta g_{0i}=aG_i$, and $\delta g_{ij}=a^2(C_{i,j}+C_{j,i})$, with
transverse conditions $G_{i,i}=C_{i,i}=0$, or, in our setup,
$G_z=C_z=0$. Then, we impose the gauge condition $\delta
B=0$. However, this condition alone does not completely fix the gauge,
as only the component of $\xi^i$ parallel to $\vec b$ gets frozen by
Eq.\ (\ref{eq:deltB}). Therefore we can still choose $\xi^y$ such that
$C_y=0$, and $\vec C=\vec C^\parallel$ is parallel to $\vec
b$. Finally, we find that the action for the vector perturbations is
given by
\begin{align}
  \label{eq:VC1}
  S_V&=\int\de^4 x \left\{ \frac{a}{32\pi G} \bigl[ {\left( \partial_z V_x \right)}^2+{\left( \partial_z V_y \right)}^2 \bigr]+na^3 b\delta A {\dot C_x} \right.\notag \\ 
&\qquad\left.{} + na^2 b V_x \delta A+2\pi Gb^2n^2a\delta A^2/\dot H \right\},
\end{align}
where $V_i \equiv G_i-a {\dot C_i}$ is a gauge invariant field. This
action can be immediately extended to the general case where the
perturbation variables depend now on $(x,y,z)$.  In the gauge $\delta
B=0$, the result is given by
\begin{align}
\label{eq:VC2}
  S_V&=\int\de^4 x \left[ \frac{a}{32\pi G} \left( \partial_j V_i \right)\left( \partial_j V_i \right)+a^3(\rho+p) {\dot C_i} \delta u_i \right.\notag \\ 
&\qquad\left.{} +a^2 (\rho+p) V_i \delta u_i-\tfrac{1}{2}\,a\,(\rho+p)\delta u_i \delta u_i \right],
\end{align}
where we substituted $\delta u_i$ for $b_i\delta A/\mu$. Variations
with respect to $V_i$ and $C_i$ yield the following equations,
\begin{align}
\label{eq:vectN}
\triangle V_i&=16\pi G a(\rho+p) \delta u_i, \\
\frac{\de}{\de t}[(\rho&+p)a^3\delta u_i]=0,
\label{eq:vectT}
\end{align}
respectively. Again, these equations exactly coincide with those
derived by perturbing the Einstein equations and the energy-momentum
conservation law \cite{Weinberg}. This provides thus a cross-check that the calculations presented here
are correct. In fact, the main novelty in our approach is to be found when we quantize the system.

To summarize section III, the known results on first-order TCP for a
perfect fluid can be directly derived from variations of the classical
action given in Eq.\ (\ref{eq:act1}). Note that a similar action
approach has been already performed in \cite{Garriga,Boubekeur:2008kn}. 
Yet, the system studied there cannot represent a perfect fluid. 
Indeed, as already mentioned in the introduction, the action proposed in \cite{Garriga,Boubekeur:2008kn} 
is made of a real scalar field. So, this system, by construction, cannot have vector perturbations, 
as the only new perturbed field is the scalar one. Therefore the system
studied there is not a perfect fluid, otherwise a perfect fluid would have no vector perturbation. 
It can be thought of as a scalar fluid, but, once more, not as a perfect fluid. 
It is simply a different physical system whose squared sound speed $c_s^2$ is not equal to $\dot p/\dot\rho$.
  %Even mathematically this was clear from the
  %beginning because for a perfect fluid the 4-velocity $\delta
  %u_\alpha$ perturbations have more degrees of freedom than
  %$\delta(\partial_\alpha \phi)$, where $\phi$ is the $K$-essence
  %field.  Indeed, since four-velocity $u_\nu$ is merely expressed in
%terms of a single scalar, the perturbations produced are only of
%scalar type: this is the reason we would call it scalar
%fluid.

\section{Quantization}
The most important advantage of the action approach proposed in this
paper is, of course, that it allows us to quantize the system. Although the
inhomogeneities of the present universe, such as the galaxy
distribution, are clearly described by the classical theory, the
quantization of a perfect fluid may have something to do with the
early universe if the seeds for structure formation are provided by
quantum fluctuations of fields generated during inflation.  Yet,
besides its practical utility, our action approach also opens new
theoretical prospects, as discussed below.  In the following, we will
again treat the quantization for the scalar and vector type
perturbations separately.

\subsection{Scalar type perturbations}
To quantize the scalar perturbations, let us first introduce the
canonical field $\psi\equiv\sqrt{a^3 Q_S} \Phi$.  To avoid the
appearance of a ghost, we assume that $Q_S$ is positive.  According to
Eq.\ (\ref{eq:Qs}), this means that $(\rho+p)/c_s^2 >0$. Such a
constraint, together with the stability of the perturbations,
$c^2_s>0$, lead to the null energy condition $\rho+p>0$. Using the
new variable $\psi$, the action (\ref{eq:act2}) is rewritten as
\begin{align}
\label{eq:act2.5}
  S_S=\int \de^4x&\left[ \frac{\dot\psi^2}2
-\frac {c_s^2}{2a^2}(\vec\nabla\psi)^2
+C_1 \delta s\dot\psi+C_2 \delta s \psi\right.\notag\\
&-\left.\frac{N}2(\delta\bar\theta\dot{\delta s}
-\delta s\dot{\delta\bar\theta})-\frac D2 \delta s^2\right] ,
\end{align}
where we have neglected $\delta A$ and $\delta B$ as they do not contribute
to the Hamiltonian.  The field $\psi$ has a canonical kinetic term,
whereas the quadratic terms for $\delta s$ and $\delta \bar\theta$ are
at most linear in their time derivatives.  Yet, it is known
\cite{Jackiw} that a consistent quantization of such a singular
Lagrangian can be done provided one introduces the following
equal-time commutation conditions,
\begin{align}
  \bigl[\hat{\psi}(t,\vec x),\hat{\pi}(t,\vec y)\bigr]&=i\delta(\vec x-\vec y)\, ,\label{comm1}\\
  \bigl[\hat{\delta s}(t,\vec x),\hat{\delta\bar\theta}(t,\vec y)\bigr]&=-\frac{i}{N}\delta(\vec x-\vec y)\, . \label{comm2}
\end{align}
All the other commutators are zero and $\pi$ is the canonical
conjugate momentum of $\psi$.  The corresponding Hamiltonian is given
by
\begin{align}
\label{eq:hamiltonian}
  {\hat H}=\int \de^3\vec x&\left[\frac12\,{\left({\hat \pi}-C_1 \hat{\delta s} \right)}^2+
\frac {c_s^2}{2a^2}(\vec\nabla {\hat \psi})^2\right.\notag\\
&-\left.C_2 \hat{\delta s} {\hat \psi}
+\frac{D}{2} {\hat {\delta s}}^2 \right].
\end{align}
One can easily check that the Heisenberg equations, with the help of
the commutation relations, yield the same equations of motion as the
classical ones derived from the variation of Eq.~(\ref{eq:act2.5}).

The relation (\ref{comm2}) shows that $\hat{\delta s}$ and
$\hat{\delta\bar\theta}$ become non-commuting variables at the quantum level. 
In Quantum Field Theory, different fields (i.e., different particles) can be simultaneously
observed at the same position. 
Here the perturbation fields related to the entropy and the temperature, 
to which we may individually attribute arbitrary numbers at the classical level, 
cannot be measured at the same space-time point. 
That this non-commutativity arises from the action of a perfect fluid is thus intriguing. 

We should concede that consequences directly linked to present observations are missing.
However, at this level it is quite interesting to compare the action (\ref{eq:act2.5}) with
the one of the Landau problem \cite{Jackiw}, an archetype of non-commutative
geometry. Regarding $\hat{\delta s}$ and $\hat{\delta\bar\theta}$, the
action (\ref{eq:act2.5}) is essentially the same as the one for a
charged particle moving on a two-dimensional surface with a constant
magnetic field background in the transverse direction:
\begin{equation}
  \label{eq:LPR}
  S=\int \de t\left[\frac m2(\dot x^2+\dot y^2)-\frac{\cal B}2(\dot x y-\dot y x)-V(x,y)\right].
\end{equation}
Within this analogy, the perturbation fields $(\delta s, \delta \bar
\theta)$ correspond to the $(x,y)$ space coordinates for the particle,
and the number of particles $N=na^3$ plays the role of the constant
magnetic field ${\cal B}$. Interestingly enough, while the quite heuristic
non-commutative relation $[\hat x,\hat y]=-i/{\cal B}$ in the Landau
problem \cite{Jackiw} holds only in the absence of the kinetic term in
Eq.\ (\ref{eq:LPR}), which is valid in the large magnetic field limit,
the non-commutative relation (\ref{comm2}) of a perfect fluid is exact
for any finite number of particles.  So, perfect fluids provide a nice
example of non-commutativity between different fields.

The other non-commutation relation (\ref{comm1}) leads also to an
interesting physical consequence. By using once more the Einstein
equations and the energy-momentum conservation law, we find that
the pressure perturbation in the comoving gauge ($v=0$) is given by
$\hat{\delta p}=-(\rho+p){\dot {\hat \phi}}/H$. Then, the commutator
between $\phi$ and $\delta p$ becomes
\begin{equation}
 \bigl[\hat{\phi}(t,\vec x),\hat{\delta p}(t,\vec y)\bigr]=-i c_s^2 H\delta(\vec x-\vec y)/a^3. 
\end{equation}
Consequently, local curvature and pressure perturbations cannot be measured simultaneously.

\subsection{Vector type perturbations}

Time derivatives of $V_i$ and $\delta u_i$ do not appear in the action
(\ref{eq:VC2}). Therefore, those are auxiliary fields which can be
eliminated through their equations of motion. The action
(\ref{eq:VC2}) becomes then a functional which depends only on $C_i$.
To make this action canonical, we introduce a new variable $F_i ({\vec
  k},t)=\sqrt{a^3 Q_V (k,t)} C^{\parallel}_i({\vec k},t)$, where
$C^\parallel_i({\vec k},t)$ is the Fourier transform of $C^\parallel_i
({\vec x},t)$ and $Q_V$ is given by
\begin{equation}
Q_V (k,t)=\frac{a^2 k^2 (\rho+p)}{k^2+16\pi G a^2 (\rho+p)}.
\end{equation}
To avoid the appearance of ghosts, $Q_V$ must be positive.  So, as for
the scalar modes we require $\rho+p>0$, i.e.\ the null energy
condition to hold.  In terms of $F_i$, the canonical action in Fourier
space is given by
%\begin{align}
\begin{equation}
S_V= \int \de t \de^3k \, \left( \tfrac{1}{2} \dot{F_i^\ast} \dot{F_i} -\tfrac{1}{2} m_k^2 F_i^\ast F_i \right), \\
\end{equation}
with
\begin{equation}
m_k^2=-\frac12\frac{\de^2}{\de t^2} \log (a^3 Q_V)-\frac{1}{4} {\left( \frac{\de}{\de t} \log a^3 Q_V \right)}^2.
\end{equation}
%\end{align}
Now the quantization is done by imposing the following canonical
condition for $F_i$ and its conjugate momentum
% $\pi_i^{\dagger}$ as
\begin{equation}
\bigl[\hat{F_i}(t,\vec k),{\hat{\pi_j}}^\dagger (t,\vec k')\bigr]=i \delta(\vec k-\vec k') \left( \delta_{ij}-\frac{k_i k_j}{k^2} \right). \label{comvec}
\end{equation}
The corresponding Hamiltonian is given by
\begin{equation}
{\hat H}= \int \de^3k \left( \tfrac{1}{2} {\hat \pi_i}^\dagger {\hat \pi_i} +\tfrac{1}{2} m_k^2 {\hat F_i}^\dagger \hat F_i \right),
\end{equation}
and the evolution of the operators is given by the Heisenberg equation
with the help of the commutation relation (\ref{comvec}). The quantum version of Eq.\ (\ref{eq:vectN}) implies $\bigl[\hat{V_i}(t,\vec x),\hat{\delta u_j}(t,\vec y)\bigr]=0$.
Therefore, the gauge invariant metric perturbation and the vorticity of the perfect fluid
can be measured at the same time, at the same position.

As for the tensor perturbations, they come only from the metric
perturbation.  The action for the tensor perturbations and its quantum
aspects have been widely studied in the literature (e.g. \cite{Weinberg}), mainly in
connection with the quantum generation during inflation. So, we do not
discuss it any longer.

\section{Conclusions}
We have studied the theory of cosmological perturbations for a perfect fluid in GR
at the action level.
Starting from the action proposed by Schutz, we first reproduced the known results derived from 
the equations of motion alone.
This enabled us to illustrate the advantage of our action approach at the classical level.
Quantizing then the perturbation fields, we found that some of
them do not commute, leading thus to a non-commutative
field-geometry. In particular, we pointed out that a simultaneous measurement of local curvature 
perturbations and pressure inhomogeneities is not allowed at the quantum level. 
Finally, we proved that both the null energy condition and a positive squared sound speed
%$c_s^2=\dot p/\dot \rho$ 
have to hold at all times in order to avoid ghost degrees of freedom.

Another advantage of our action approach is that one can easily obtain the second order action 
depending only on the dynamical fields.
Such an approach is thus suited to study cosmology in extended gravity models with more than one 
dynamical field.
In particular, we expect that the approach presented here will be quite useful for the perturbation analysis 
of $f(R,G)$ gravity models \cite{ANTONIO}, or for the treatment of non-gaussianities for the entropy and vector perturbations on perfect fluids following \cite{Boubekeur:2008kn}.

%{\bf These results will be hopefully important in our present picture of cosmology. They should help providing the starting point for more complicated studies, such as imprints %of quantum perturbations on cosmological observables. For examples non-gaussianities produced during or just afer reheating, if modeled by a perfect fluid should be studied via %the approach introduced in this paper. We will consider this interesting topic as a project to be addressed in the future.}

\begin{acknowledgments}
  We thank Sean Murray for helpful discussions. This work is supported
  by the Belgian Federal Office for Scientific, Technical and Cultural
  Affairs through the Interuniversity Attraction Pole P6/11.
\end{acknowledgments}

\end{document}